\documentclass[apj]{emulateapj} \setkeys{Gin}{width=0.48\textwidth}
\usepackage{graphicx}

\newcommand{\hmpc}{\ifmmode{h^{-1}\,\hbox{Mpc}}\else{$h^{-1}$\thinspace Mpc}\fi}

\newcommand{\kms}{\ifmmode{\,\hbox{km\,s}^{-1}}\else {\rm\,km\,s$^{-1}$}\fi}
\newcommand{\msun}{{\rm\,M_\odot}}

\begin{document}
\title{Star Streams and the Assembly History of the Galaxy} %\altaffilmark{1}}
\shorttitle{Star Streams  and Galaxy Assembly}
\shortauthors{Carlberg}
\author{Raymond G. Carlberg}
\affil{Department of Astronomy \& Astrophysics, University of Toronto, Toronto, ON M5S 3H4, Canada} 
\email{carlberg@astro.utoronto.ca }

\begin{abstract}
Thin halo star streams originate from the evaporation of globular clusters and therefore provide
information about the early epoch globular cluster population. 
The observed tidal tails from halo globular clusters in the Milky Way are 
much more shorter than expected from a star cluster orbiting for 10~Gyr.  
The discrepancy is likely the result of the assumption that the clusters have been orbiting in a
non-evolving galactic halo for a Hubble time. 
As a first step towards more realistic stream histories, a toy model that combines an idealized merger model
with a simplified model of the internal collisional relaxation of individual star clusters is developed.
On the average,   the velocity dispersion
increases with distance causing the density of the stream to decline
with distance. Consequently, 
the streams visible in current data will normally be some fraction of the entire stream. 
Nevertheless, the high surface density segment of the stellar streams created from the evaporation of the 
more massive globular clusters should all be visible in low obscuration parts of the sky if closer than about 30~kpc.
The  Pan-STARRS1 halo volume is used to compare the numbers of halo streams and globular clusters.
\end{abstract}

\section{INTRODUCTION}
\nobreak
The current globular cluster population is the remnant of a larger population which lost its lower mass
members to evaporation driven by internal two-body relaxation
and tides \citep{FR:77,GO:97,FZ:01}. However, at least some fraction of those
remnants are visible as the thin
 stellar streams found in the halo \citep{GC:16}. Combining 
globular clusters and their remnant stellar streams into one
dynamical picture offers the possibility of providing additional, less ``survivor" biased, insights
into the early epoch globular cluster population and the assembly history of the Milky Way halo.

Two-body relaxation in globular clusters causes stars outside the core to gain
energy, which drives a gradual evaporation of the cluster in the tidal field of the galaxy.
The evaporating stars form tidal star streams
that are useful for  measuring the shape
of the galactic potential \citep{Binney:08,LM:10,EB:11,SB:13,Bovy:16} 
and detecting the many dark sub-halos that are expected to be present \citep{Klypin:99,Moore:99}.
The dynamical analysis of stream formation usually starts with
star clusters instantaneously inserted into a fixed background potential, although that potential may contain 
a small fraction of its mass in the form of orbiting sub-halos.  
Some simulations use an approximation that prescribes the release of stars near the tidal surface
\citep{Kupper:12}.
In n-body simulations of stream formation  particle softening makes the clusters collisionless,
although tidal heating in a reasonably elliptical
orbit 
is sufficient to cause mass loss. 
Although the mass loss rate varies around the orbit, it is periodic so that 
the the mean mass loss rate is constant and the
dynamical properties of the stars along the stream repeat nearly exactly along the stream
\citep{DOGR:04,Carlberg:15a}. Much valuable insight results from the analysis of streams and
sub-halos orbiting in a non-evolving potential \citep{Kupper:12,Carlberg:15a,Bovy:17} but it does
overlook the fact that the overall halo potential and is assembled over comparable
time-scales. 

Observed stellar streams, or stream segments, are typically about 10 kpc such as the well-studied  Pal~5 
\citep{Odenkirchen:01,GD:06Pal5} and GD-1  \citep{GD:06GD} streams.
There length is somewhat
puzzling because globular clusters are 10 Gyr old systems, that release stars into the tidal stream
at typical velocities, about 5 \kms, which should
lead to long streams, $\sim 50$ kpc. 
How the process of assembly of 
the globular clusters into our galactic halo affects the properties of the stellar streams, and,
conversely,  how stellar streams  might provide insight into the history of the halo and its globular cluster population
is the goal of this paper.

The internal dynamical evolution of globular star clusters in a tidal field is fairly well
understood \citep{Spitzer:87,Davies:13}. 
The mass evolution of globular clusters in the
galaxy halo has been discussed at length  \citep{FR:77,FR:85,GO:97,FZ:01,GHZ:11,LBG:13}.
\citet{RGB:11} have
developed an  approach to the problem of a star cluster evaporating in a merging system, concentrating 
on the star cluster evolution.
The rate of evaporation depends
on the internal  two body relation time and the local tidal field. The tidal field changes
from the site of cluster formation to its accretion into the galactic halo.
As a cluster evaporates the rate and velocity at which stars are
ejected into the tidal streams changes,  which in turn affects the density of the tidal
stream created. 

This paper explores how 
the observed star streams and remnant halo globular clusters  
are related with their history of assembly into the galaxy.
A toy model of accretion infall of a satellite galaxy containing model globular clusters is developed.
The satellite is represented as an evolving analytic potential which falls into the host with dynamical friction.
The particles in the model clusters are subjected to a simple Monte Carlo heating procedure which mimics
the relaxation processes of globular clusters with the expected mass and size scaling. 
The evolution of the clusters with time and the resulting properties of the streams on the sky are presented.
As an initial practical application, the number of star streams in the recent Pan-STARRS1 volume 
is compared to the number of 
globular clusters.

\section{A Simplified N-body Model}

An n-body simulation of  the dynamical evolution of a star cluster starting
in a satellite galaxy which merges with its host
ideally requires a large numbers of particles and a high precision n-body code
that follows stellar two-body interactions but has no two-body interactions with dark matter particles.
Here  the problem is split into two parts,
one being to model the dark matter merger of a satellite and host galaxy and the other
to provide internal evolution of the star cluster. 
The aspect of galaxy merging needed 
here is a satellite galaxy potential 
that  loses orbital energy and blends into the host
galaxy releasing its  globular clusters to orbit in the host galaxy potential.
Analytic potentials for the host and satellite combined with dynamical friction can meet these requirements.
There is no need in this exploratory study to have a full dynamical
model of the merger process or detailed cosmological starting conditions. 

The the initial host galaxy model is Milky Way 2014 of \citet{Bovy:15}. 
The halo mass component of the model potential absorbs mass from the satellite  as it merges, but there is
no violent relaxation.
A population of dark matter sub-halos is included in the host galaxy as fixed mass Hernquist spheres
following the prescriptions of \citet{CG:16}. The sub-halos extend out to 360 kpc.  
Sub-halos below a mass of $0.9\times 10^7\msun$
are not included because they make essentially no difference
to the streams studied here.

\begin{figure}
\begin{center}
\includegraphics[angle=-90,scale=0.9]{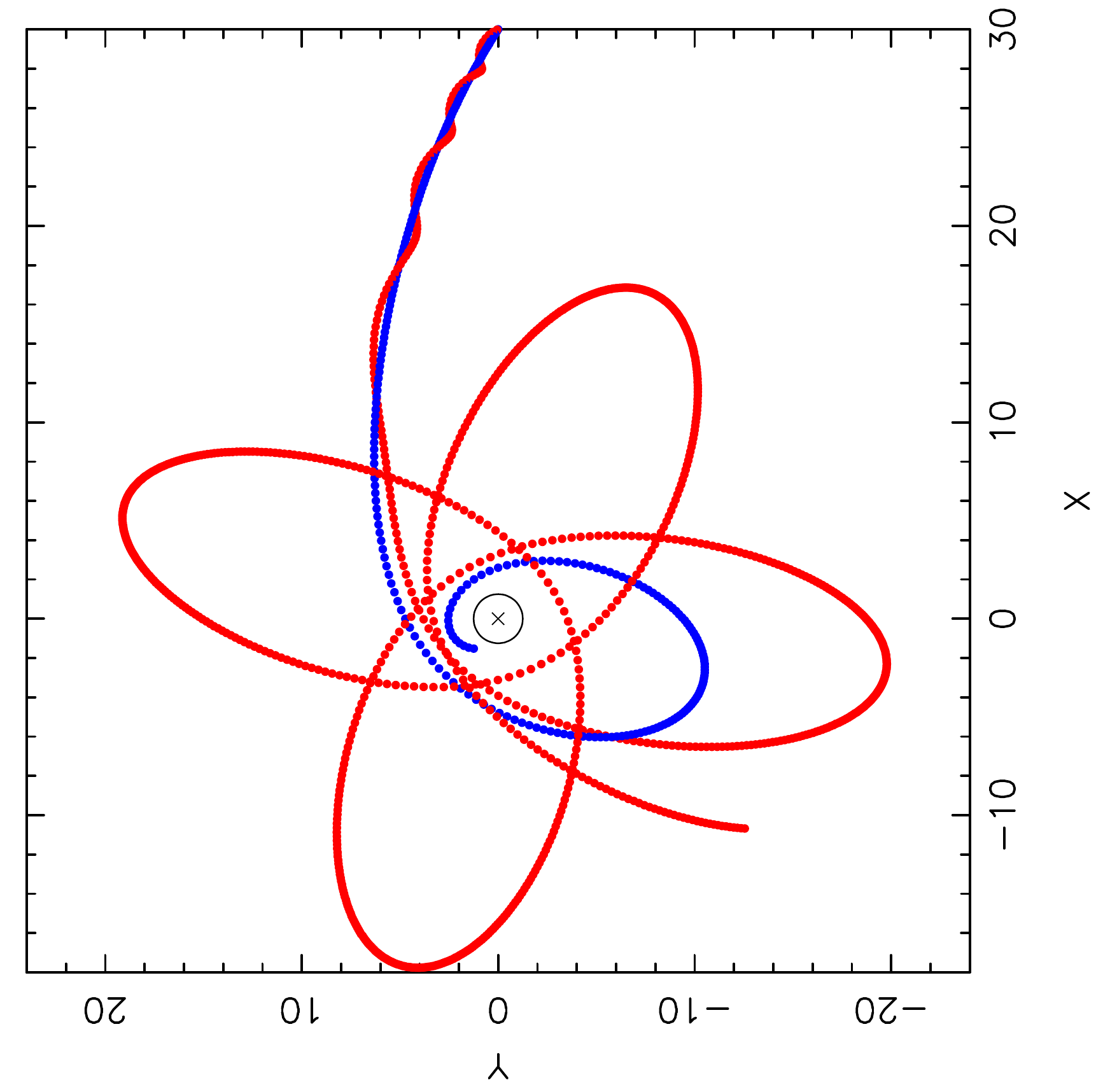}%{xymerger.pdf}
\end{center}
\caption{The blue points show the orbit
of the center of the satellite potential with $M_s=0.3$. The red points show the orbit of the center
of a single model star cluster.
The scale is in the dimensionless units of 8 kpc per unit. 
More massive satellites sink more quickly with $M_s=2$ having only a quarter turn orbit.
}
\label{fig_xy}
\end{figure}

\subsection{The Satellite Galaxy Starting Conditions}

The infalling galaxy needs to have a well-defined mass to allow the dynamical friction formula
 to be used.
The Hernquist model \citep{Hernquist:90}, $\phi_H(r,a_s)=-GM_s/(r+a_s)$  is a convenient choice 
where $M_s$ and $a_s$ are 
the satellite mass and scale radius, respectively.  
The Bovy MW2014 potential has  a halo  mass inside the 245 kpc virial radius of
$8.1\times 10^{11}\msun$ or 8.992 dimensionless mass
units, where the mass unit is $9.006\times 10^{10}\msun$. 
The mass lost is added to the host galaxy halo NFW potential \citep{NFW} which has
a scale radius of 2 dimensionless units in MW2014. The satellite needs to have a similar dark matter
density, so $a_s = 2 (M_s/8)^{1/3}$.
The baseline satellite mass,  $M_s$, is  chosen to be 2 mass units  with $a_s=1.26$ which can be
considered as a satellite with 22\% of the mass of the host halo within the virial radius.

Ideally, the satellite galaxy orbit would be drawn from a cosmological simulation, but with a single infalling galaxy all 
that is required is a reasonably representative starting point someplace near the virial radius.
The satellite galaxy is started in a bound orbit, at a galactocentric $x$ of 240 kpc and above the plane of
the MW2014 model at $z$ of 120 kpc. The satellite is given an initial tangential velocity equal to 0.35 of the
local circular velocity and an inward radial velocity of 0.61 of the local circular velocity. 

\subsection{The Merger Model}

An accurate violent relaxation model
is not essential for this investigation which focuses on the star clusters and their tidal streams.
Two analytic time-varying analytic potentials moving together with dynamical friction give
an approximate description of the merger. 
Chandrasekhar's formula \citep{BT:08,BoylanKolchin:08} is used to evaluate the dynamical 
friction, with $\ln{\Lambda}=10$ adopted. 
The NFW dark matter halo in the MW2014 potential is used to calculate the local mass density 
for friction. The NFW is given a small
core radius equal to the MW2014 bulge core radius to avoid the singularity when the satellite is
dragged to the center of the host galaxy.
As the satellite merges into the host, 
it loses mass and expands to spread over the host halo potential.   
The key dynamical quantity is $\Omega_s$, the local circular frequency of the satellite orbit, which is
 estimated  from the local acceleration 
in the host galaxy, $\Omega_s= \sqrt{\bf a\cdot r}/r$.
The spread in circular frequency of the orbit at radius $r$ across the scale radius of the satellite,  $a_s$, 
is approximated $\delta \Omega_s\simeq \Omega_s a_s/r$.
 Therefore, the rate of satellite mass decrease due to orbital divergence is 
modeled as,
\begin{equation}
{dM_s \over dt }= - {M_s a_s\over r^2}\sqrt{\bf a\cdot r},
\label{eq_mt}
\end{equation}
with the lost mass being added to the host halo.
The scale radius of the satellite, $a_s$, expands as 
\begin{equation}
{da_s\over dt} = { a_s\over r}\sqrt{\bf a\cdot r}.
\label{eq_at}
\end{equation}
These ideas are implemented in the simulation, with 
the maximum scale radius of the satellite  limited to 5 times the scale radius of the NFW of the host galaxy.
The mass taken from the satellite is added to the host galaxy halo without changing the  scale radius. 
Under this process
the satellite orbits into the center of the host galaxy, expands and its mass goes to zero.
An example merger orbit is shown in Figure~\ref{fig_xy}.

\begin{figure}
\begin{center}
\includegraphics[angle=-90,scale=0.7,trim=80 60 50 80]{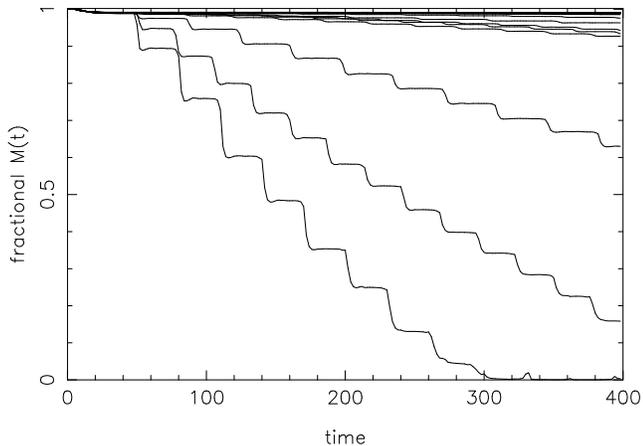}%{evaporate825.pdf}
\end{center}
\caption{Mass as function of time for the 24 clusters each with mass $0.9\times 10^4 \msun$
with only tidal heating. The internal relaxation model is turned off for the purpose of comparison.
}
\label{fig_mt825}
\end{figure}

\subsection{Initial Conditions and Gravity Code}

Star formation  occurs in galactic gas disks, which includes the dense star systems
that are the likely current epoch equivalents of globular cluster progenitors.  Observationally, the formation of 
globular clusters may be enhanced in vigorously star forming systems
associated with interacting disk galaxies \citep{Whitmore:99}. 
The numerical modeling of 
the formation of progenitors to globular clusters
\citep{KG:05,BDE:08,PZ:10,RAG:16} is being developed.  The simulations find 
a steep cluster mass distribution, similar to that observed in the dense stellar systems, nearly $M^{-2}$. 
The simulations here follow a single cluster
mass at a time, but simulations at different masses can be combined with
appropriate weights to determine how a mass distribution will evolve.

The star clusters start as King models with a  concentration parameter of $W_0 =4$. There are 
50,000 particles in a single star cluster. 
A model cluster of mass $M_c$  at an orbital radius $r_o$
in the satellite is given an outer radius scaled to the local tidal radius,
$r_t = (M_c/3 M_t)^{1/3} r_o$, with the 
 tidal mass 
$GM_t = r\,{\bf a\cdot r}$.
The acceleration is calculated from the satellite galaxy potential alone. 
The cluster particles are evolved with a fast parallel  shell gravity code which accurately reproduces the results
of a full n-body code \citep{Carlberg:15b} for these very low mass clusters relative
to the host galaxy. The shell code 
calculates self-gravity using only the gravitational monopole, that is,
the gravitational acceleration at radius $r$ from the center of the cluster is $-GM_c(<r)/(r^2+\epsilon^2)$ 
in the radial direction,
where $M_c(<r)$ is the cluster mass interior to the location of some particle 
and the softening, $\epsilon$, is set equal to one-fifth of the core radius. T
he softening is set equal to 0.2 of the core radius of the King model,
which varies with the scaling of the model to the tidal radius. Typically $\epsilon$ is about 1~pc.

Stellar mass loss is incorporated with an analytic model.
Here, all particles have masses that decline as $m_p(t)= m_p(0)[1-(t/t_w)^\gamma$.
We use $t_w=16000$ ($\simeq 560$ Gyr)  and $\gamma=0.5$ which approximately describes the mass loss
for ages beyond the evolution over 1-14 Gyr. At the final moment of the simulations the individual
particles have lost 16\% of their initial mass. This mass change makes little difference
to the outcome.

A set of 24 star clusters is placed on circular orbits within the infalling satellite galaxy, 12 at the satellite's
scale radius $a_s$ and 12 at $\onehalf a_s$.
The clusters are in a disk that is  tilted  at 45\degr\ with respect to the plane of the host galaxy potential. 
In physical terms, the clusters begin within about 10 kpc of the center of their satellite host, which is itself at a distance
of about 250 kpc from the center of the main host galaxy.  

Figure~\ref{fig_xy} shows a typical infall path of the satellite and a single cluster. 
The cluster orbits around the center of the infalling galaxy which is losing 
mass and spreading out as it falls in. 
When the  satellite galaxy 
has fallen to a galactocentric radius of 5-10 units its mass has dropped to 
about half its initial value and the star clusters are unbound from the dissolving satellite galaxy. 
The satellite galaxy in
Figure~\ref{fig_xy} has $M_s=0.3$. A satellite with $M_s=2$ makes only a one quarter turn orbit before plunging to 
the center.

\subsection{A Simple Model of Cluster Internal Dynamics}

The King model clusters are effectively collisionless  over a model Hubble time.
Hence, with only weak tidal heating,  evaporation of star particles  will be very slow, 
with even the most eccentric orbits losing less than 10\% per Hubble time,
as shown in Figure~\ref{fig_mt825}. 

The internal dynamics of globular clusters is a rich subject which continues to develop with ever more realistic n-body 
simulations and sophisticated dynamical analysis. 
The dynamical evolution of star clusters is reviewed in \citep{Spitzer:87, BT:08,Davies:13}. 
In brief, two body relaxation leads to the core shrinking and envelope expansion
until heating from central binary and multiple star systems intervenes to 
limit core collapse and sometimes cause core oscillations. 
The gravitational collisions, largely in the central region, cause stars in the envelope
to gradually move to larger orbits until they reach a tidal surface and become unbound from the cluster. 
\citet{GHZ:11} find that clusters with mass greater than $(8~{\rm kpc}/R)4\times 10^4 \msun$, 
are in the expansion dominated
phase of evolution, where $R$ is the galactocentric radius of the cluster.  
The vast majority of thin Milky Way streams are located outside the solar circle and well beyond.
A typical halo cluster at 15 kpc that is more
massive than $2 \times10^4\msun$ will be in the expansion phase of evolution, which simplifies
the modeling requirements.

\begin{figure}
\begin{center}
\includegraphics[keepaspectratio,scale=0.8,trim=130 180 160 190]{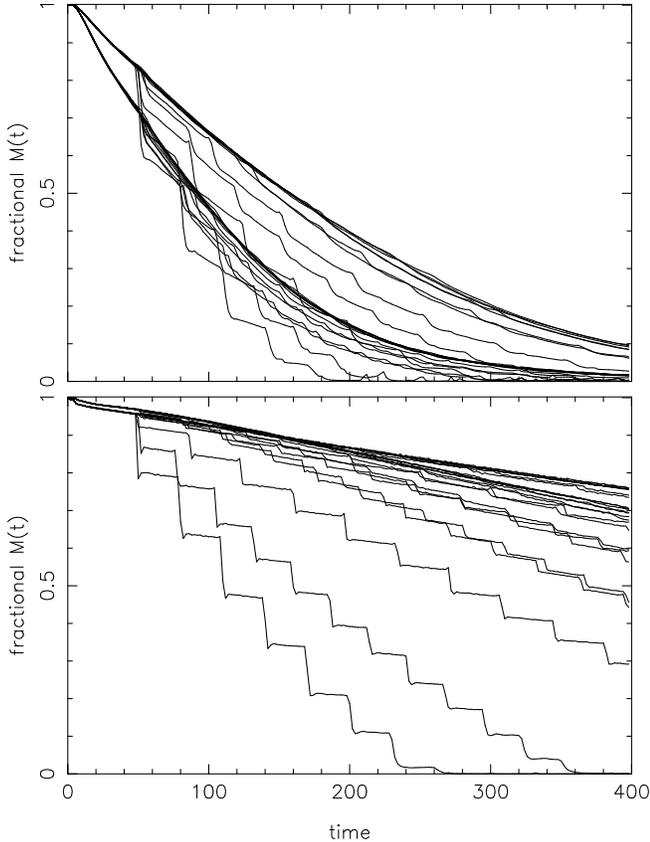}%{MT.pdf}}
\end{center}
\caption{Mass as function of time for the 24 clusters  each initially with mass $0.9\times 10^4 \msun$  
in the top panel and $0.9\times 10^5\msun$ in the bottom panel. 
The heating model is incorporated into all of the clusters, with
the same parameters for all. Highly eccentric orbits have large mass
loss variations around the orbit.
}
\label{fig_mt45}
\end{figure}

\begin{figure}
\begin{center}
\includegraphics[keepaspectratio,scale=0.8,trim=130 180 160 190]{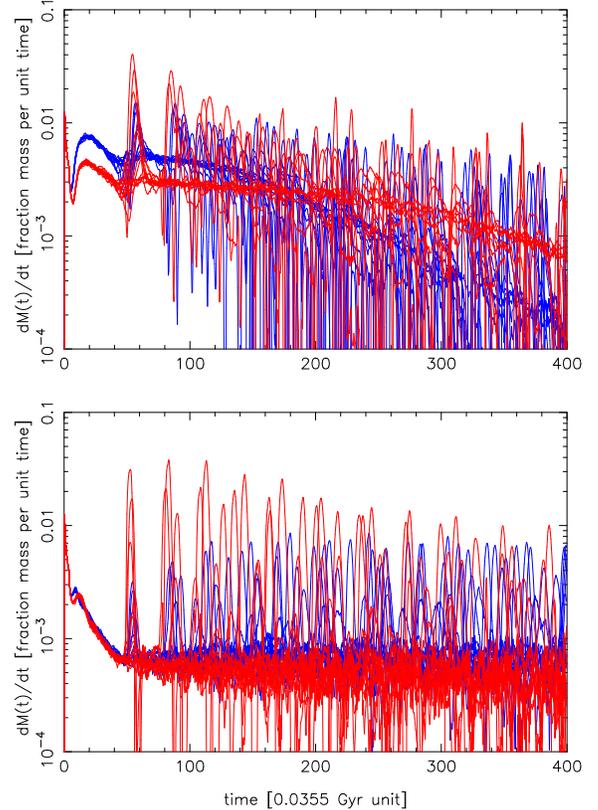}%{DMT.pdf}}
\end{center}
\caption{Mass loss rate, normalized to the initial mass, as function of time for the 24 clusters 
 each with initial mass $0.9\times 10^4 \msun$  
in the top panel and $0.9\times 10^5\msun$ in the bottom panel. 
The inner 12 clusters use blue lines and the outer 12 are in red.
}
\label{fig_dmt45}
\end{figure}

The controlling timescale for internal evolution of a  cluster of mass half mass $M_{1/2}$
 is the relaxation time at the half mass radius, $r_{1/2}$, 
\begin{equation}
 t_{rh}\propto M_{1/2}^{1/2} r_{1/2}^{3/2}\propto \sqrt{\langle v^2\rangle}  r_{1/2}^2
\label{eq_trh}
\end{equation}
 \citep{Spitzer:69, Spitzer:87}, where $\sqrt{\langle v^2\rangle}$ is a representative
half-mass velocity dispersion.
The relaxation process causes the envelops to expand in a self-similar manner
\citep{Henon:61,Freitag:06a,Freitag:06b}.

The goal here is to  understand  how
a somewhat more complete cluster and galaxy assembly history affects the stream properties.
The heating process is simply modeled with the addition of small velocity kicks to stars in the clusters. 
The required size of the velocity increments is estimated from the diffusion equation for
stellar velocities \citep{BT:08} in a cluster with collisions,
\begin{equation}
{{d  \langle v^2 \rangle } \over {dt}} = \zeta {{  \langle v^2 \rangle } \over {t_{rh}(M_{1/2},r_{1/2})}},
\label{eq_heat}
\end{equation}
where \citet{GHZ:11} find that $\zeta \simeq 0.1$.
In  a dynamically correct 
Monte Carlo simulation an energy increase to outer stars would be extracted from inner stars
and binaries \citep{Giersz:98}. 
This initial investigation does not require an accurate internal model of the cluster, so
energy is simply added to the envelope at the predicted rate.

The velocity increments, $\delta v$, will be applied at discrete intervals,
$\delta t_r$. Over some simulation time  $T$, the velocity gains are 
$\Delta \langle v^2 \rangle = T/{\delta t} (\delta v)^2$.
Comparing Equations~\ref{eq_trh} and  \ref{eq_heat} gives the required step-wise velocity changes,
\begin{equation}
(\delta v)^2 \propto \zeta {{\delta t}\over T}  {M_{1/2}^{1/2}\over r_{1/2}^{5/2}}.
\label{eq_hv}
\end{equation}
A practical implementation of Equation~\ref{eq_hv} is every $\delta t$ in time to add
random velocities in all three directions drawn from a Gaussian velocity distribution of width, 
\begin{equation}
\delta v = C    \sigma \left({\delta t\over T} \right)^{1/2}  \left({M_{1/2}\over 10^{-7}}\right) ^{1/4} 
\left({r_{1/2}\over 10^{-3}}\right)^{-5/4},
\label{eq_hc}
\end{equation}
in the dimensionless units of the calculation, 
where C is a constant around unity that absorbs $\zeta$ and $\sigma$ is a characteristic velocity dispersion in the cluster.
The $\delta v$ value is used to generate  random velocity offsets in the $xyz$ directions  from a Gaussian distribution.
In this paper $\delta v$ is fixed for any model cluster, however, 
it could be normalized to the evolving RMS velocity dispersion of the cluster. 

A little experimentation is required to fine tune the parameters controlling the
heating process. An outer radius for the evaluation of the RMS velocity and the half mass radii of the clusters is 
selected to $r_m=0.02$ units (160 pc) which is somewhat larger than the initial tidal radius
of the clusters.  
No heating is done for particles beyond $r_m$ or inside the radius containing 1/3 of the mass inside $r_m$,
the latter to roughly replicate the results of detailed collisional n-body models.
The heating calculation is done every $\delta t$ of 100 steps 
($0.71\times 10^7$ yr). Equation~\ref{eq_hc} is evaluated with $C \sigma \sqrt{\delta t/T} $ set equal
to $3\times 10^{-4}$ for $T=400$, the duration of the simulation. 
The simulations have 200,000 time steps so this random walk process happens
2000 times. 

\begin{figure}
\begin{center}
\includegraphics[keepaspectratio,scale=0.8,trim=130 180 160 190]{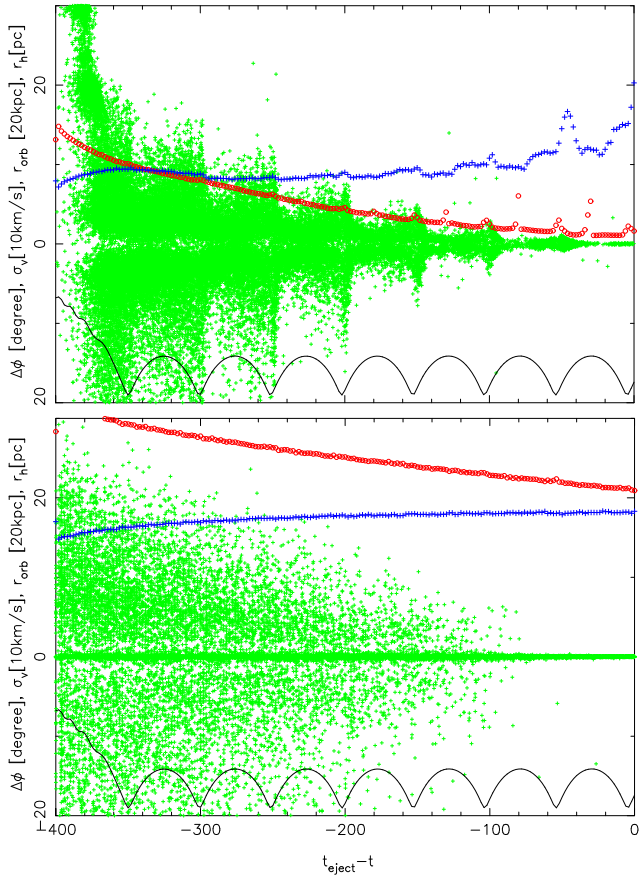}%{DLP.pdf}}
\end{center}
\caption{Evolution of the satellite and particles released into the stream with time for single clusters on the
same orbit with mass $0.9\times 10^4$ and $0.9\times 10^5\msun$ 
in the top, middle and bottom panels, respectively.
Red circles are the 3D velocity dispersion in $\kms$ times 10. 
Blue plus signs are the half mass radius of the remaining star cluster in parsecs.
Green is the angular distance on the sky between the satellite center and the released particles, 
measured in degrees.
The black line is the dimensionless orbital radius in 20 kpc units, offset by -20 to the bottom of the plot.
The orbital eccentricity is 0.86.
}
\label{fig_rspt45}
\end{figure}
 
\section{Cluster and Stream Evolution}

The simulations have 24 star clusters arranged in two rings of 12 clusters
in the satellite galaxy.  Simulations with all clusters having an initial  mass
of $0.9\times 10^4 \msun$ and $0.9\times 10^5\msun$ are done.  
The simulations reported use a satellite galaxy
with $M_s=2$ which can be considered a 22\% mass satellite, relative to the MW2014 model to its
nominal virial radius. The simulations run for 400 time units, or 14.2 Gyr.
All simulations use the same heating parameters and merger model. Simulations have been run 
for many other satellite masses, satellite starting conditions, sub-halo contents of the host galaxy and the
cluster internal relaxation parameters, however the set presented provides a good guide to understanding 
the general outcomes. 

Figure~\ref{fig_mt45} shows the fractional cluster mass as a function of time for 
initial cluster masses of $\sim 10^4 \msun$ (top panel) and $\sim 10^5\msun$ (bottom panel).  
As expected the $\sim 10^4\msun$ clusters  largely, but not completely, evaporate in a Hubble time whereas
the $\sim 10^5\msun$ clusters, on the average, lose less than half their mass. 
The mass loss rates are shown in Figure~\ref{fig_dmt45} with 
the clusters separated into the inner ring (blue) and the outer ring (red).  The peaks
in the mass loss rates with time occur at orbital apocenter, 
where stars that were unbound earlier move away from the cluster.  
The inner clusters initially have higher mass loss rates, most clearly shown
in the upper panel of  Figure~\ref{fig_dmt45}, which is a consequence of the higher tidal fields nearer to
the center of the satellite galaxy.
The outer clusters tend to be launched into more eccentric orbits where they experience stronger tidal fields
at orbital pericenter and have higher mass loss rates. 
Two of the 24 clusters of the bottom panel have sufficiently elliptical orbits that tidal fields help drive complete evaporation. 

The $\sim 10^5 \msun$ clusters here are comparable to the direct n-body simulations for
a similar mass cluster on a similar range of orbital eccentricities \citep{Zonoozi:17}. 
All the clusters have orbital apocenters near 250 kpc. The two clusters that dissolve have pericenters of 4 and 6 kpc, 
whereas the two that drop to about 1/3 and 1/2 of their initial particle count (the masses drop a little more due stellar ) 
have pericenters of 8 and 10 kpc, respectively.

The characteristic dynamical quantities of a single  cluster and its stream are illustrated  in Figure~\ref{fig_rspt45}
as a function of time
for individual clusters at masses  $\sim 10^4$ and $\sim 10^5\msun$,  in the top and  bottom panels, respectively.
The two clusters are on essentially the same orbit, with eccentricity 0.86.
The half mass radius of the clusters, around 20 pc, is well inside the tidal radius of the clusters, 
which varies around the orbit and is typically about 100 pc.  The particle density distributions have no crisp cutoff.
The half mass radius expands modestly, approximately 50\%, over the course of the cluster evolution 
as long as the cluster retains significant mass.
The RMS velocity dispersion of the cluster stars is largely dependent on the cluster mass,
dropping to zero if the cluster evaporates completely. 
The spray of (green) points in Figure~\ref{fig_rspt45} 
shows the current angular distance from the progenitor cluster (or its remnant center) of stream particles
as a function of the time at which the star particle went through a fixed radius near the tidal surface. The decline
of velocity dispersion as the cluster evaporates will be directly reflected in the properties along the stream.
The lower mass cluster loses relatively more stars when the cluster was still in the satellite galaxy. These stars spread to
relatively large angular distances away from the from the progenitor cluster. 
The particles that evaporate late do so at such low velocity dispersions that
they remain relatively close to the cluster compared to the higher drift velocities of 
particles released earlier.  
The more massive clusters lose relatively few stars while orbiting in the satellite galaxy, 
after which the cluster maintains a fairly uniform low rate
of mass loss ejecting stars at velocities that only allows particles
to reach angular separations of 20\degr\ or so from the cluster. Stream properties are
cluster mass dependent.

\begin{figure}
\begin{center}
\includegraphics[keepaspectratio,scale=0.8,trim=120 180 150 190]{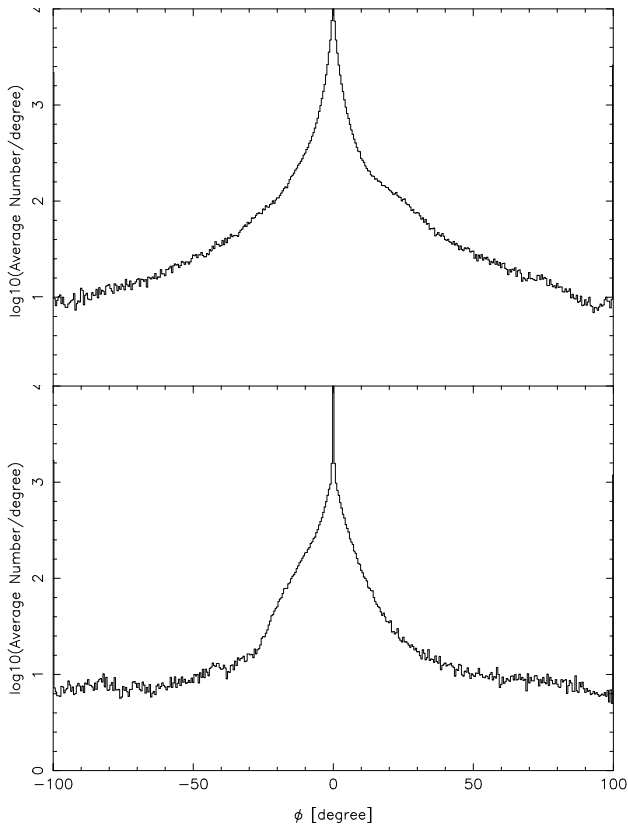}%{NPHI.pdf}}
\end{center}
\caption{The average particle density along the stream projected onto the x-y plane
averaging together  the 24 clusters with mass $\sim 10^4\msun$ in the top 
panel and $\sim 10^5\msun$  in the bottom panel, both at time 400 (14 Gyr).   A conversion
to stellar density would boost the relative density of the lower panel a factor of ten.
}
\label{fig_nphi45all}
\end{figure}

\begin{figure}
\begin{center}
\includegraphics[keepaspectratio,scale=0.8,trim=130 180 160 190]{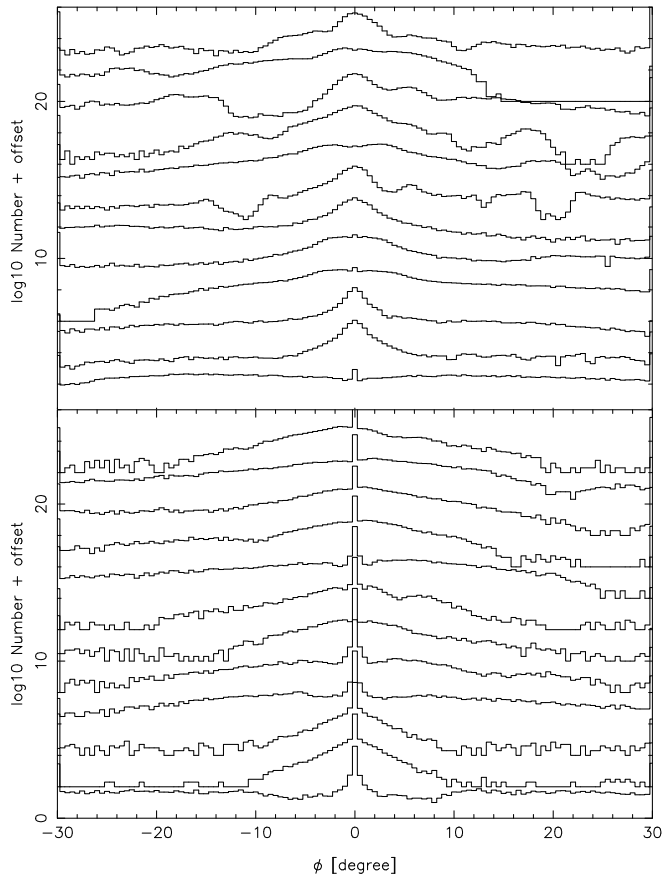}%{NPHImm.pdf}}
\end{center}
\caption{The  density along the stream for individual streams from
the 12 inner clusters with masses $0.9\times 10^4$ and $0.9\times 10^5 \msun$ 
in the top and bottom panels, respectively. Individual streams vary as a result of orbital 
phase which increases density at apocenter and decreases it at pericenter.
The 5\degr\ scale density variations along an
individual stream
are the result of dark matter sub-halos passing through the streams.
}
\label{fig_nphi45}
\end{figure}

The mean particle density summed over all  24 streams from a set of 
equal mass clusters is plotted as function of azimuthal angle with the orbit projected
onto the x-y plane in Figure~\ref{fig_nphi45all}. The average largely removes
the orbital phase variations of individual streams.
The particle density in the lower mass streams is higher because the clusters essentially
completely evaporate. 
The cluster models all have 50,000 star particles, so a conversion to
stellar density the actual stellar density would lead to a relative increase of a factor of ten for
the higher initial mass cluster streams of the bottom panel. That is, the higher mass clusters here will, on the average,
be easier to find on the sky, for the same orbital parameters. Yet higher mass clusters evaporate so slowly that
their streams are very low density. An implication is that star clusters with initial masses around 
$\sim 10^5\msun$ will produce the highest sky density star streams. 

The density along individual streams from the inner dozen clusters at
 $\sim 10^4$ and $\sim 10^5\msun$ clusters done in a simulation with sub-halos present
 is shown in Figure~\ref{fig_nphi45}.
 There is a substantial  variation in stream density profiles depending
 on the details of the orbits and the sub-halo interactions. 
The considerable variation of the stream density profiles
from one orbit to the next is clear readily visible, much of it due to orbital phase where the stream
near apocenter increases in linear density on the sky and declines near pericenter. 
Sub-halo induced gaps are
visible in most of the stream densities.  The orbits of both clusters and sub-halos are the same for the
two sets of cluster masses the gaps are nearly identical as are the sub-halo interactions, so the differences between
the two mass sets can largely be attributed to cluster mass dependence. 
 The most prominent gaps range in size from a degree up to about ten degrees,
with the most visible gaps being around 5 degrees in length.

The x-y projections of streams from $\sim 10^4\msun$ and $\sim 10^5\msun$ clusters are shown
in Figure~\ref{fig_xy45}. To a good approximation the cluster orbits are independent of the cluster masses, 
so for every cluster in the top panel there is a matching one at 10 times the initial mass in the bottom panel. 
All of these clusters are on orbits with apocenters near 240 kpc  but with a wide range of pericenters.  
The lower mass clusters on the top make narrower streams and the sky density is less concentrated 
to the location of progenitor cluster, as also shown in 
Figure~\ref{fig_nphi45all}. The satellite galaxy model potential can be adjusted to produce smaller apocenters.

\section{Visibility of Streams on the Sky}

The simulations here lead to the conclusion that streams from star clusters at masses around  $10^5\msun$ should be
the most visible on the sky, when converted to number of stars per unit angular length of stream.
The density and width varies along the stream, with the ends being relatively low linear density and relatively wide, hence
reduced surface brightness relative to near the progenitor cluster.

\begin{figure}
\begin{center}
\includegraphics[angle=0,scale=0.8]{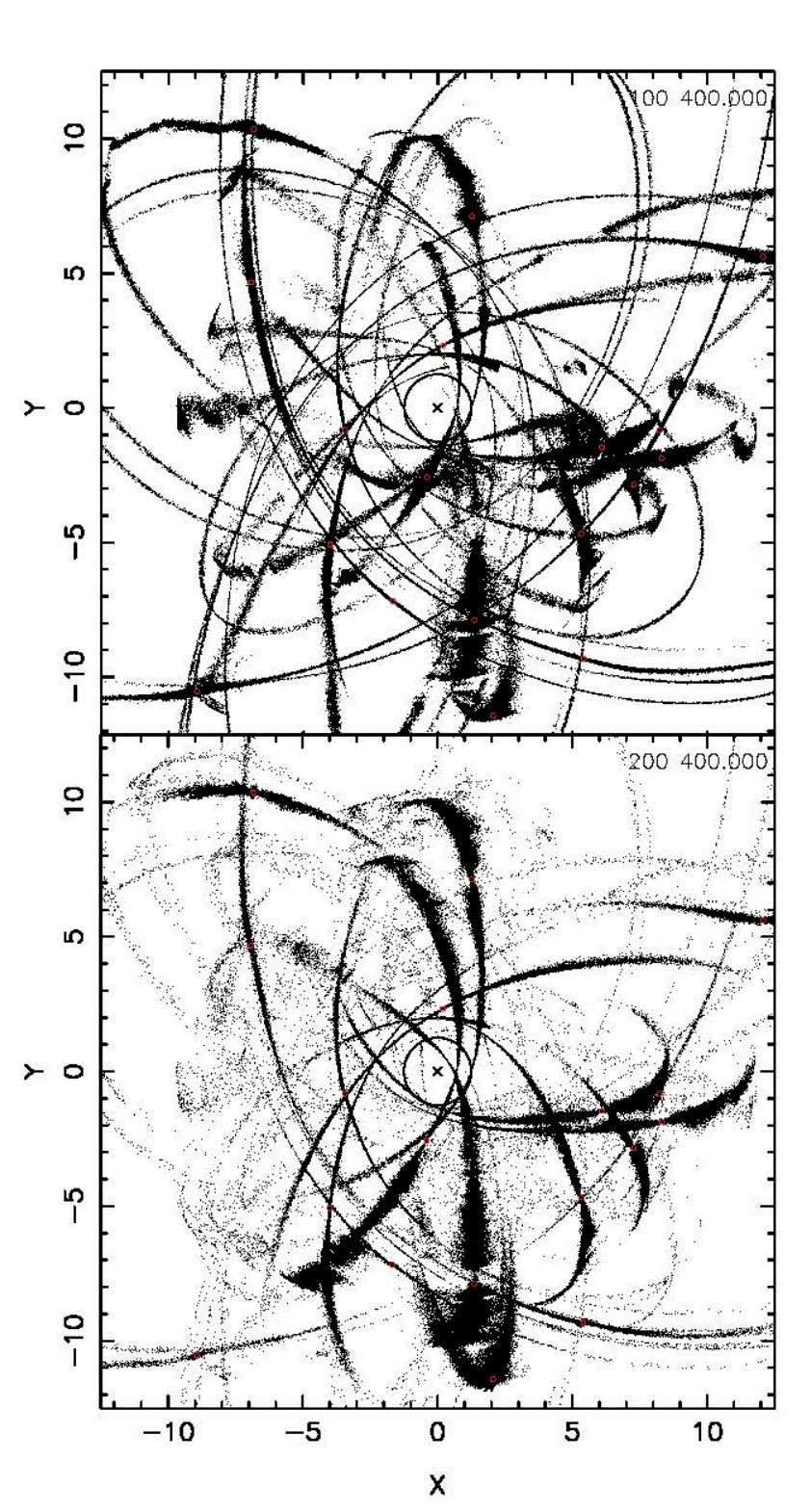}%{XY45image.pdf}
\end{center}
\caption{The tidal tails of a set of 24 infalling clusters each with mass  $\sim 10^4 \msun$ 
(top) and  $\sim 10^5 \msun$ bottom.
The grid scale is in units of 8kpc with a circle of radius 10kpc around the center.  The red circles
are the locations of the nominal cluster centers, some of which have no remnant cluster.
}
\label{fig_xy45}
\end{figure}

\begin{figure}
\begin{center}
\includegraphics[scale=0.4,trim=120 180 120 150]{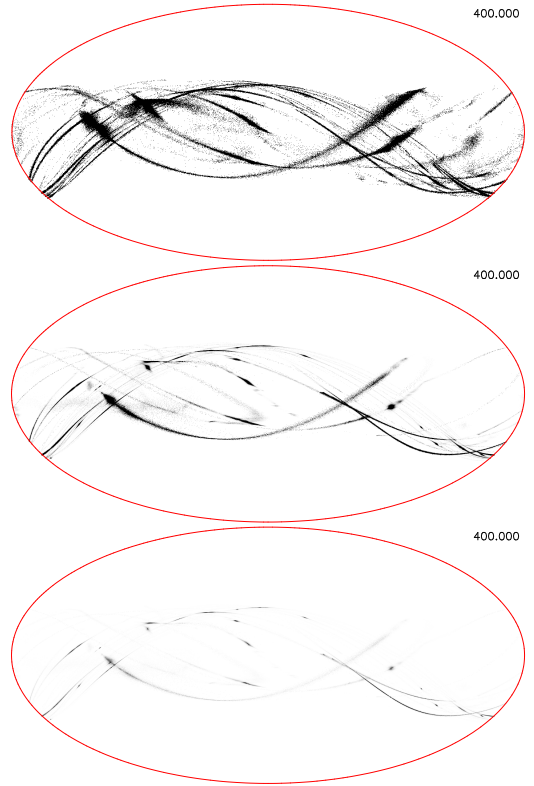}%{AH200.pdf}
\end{center}
\caption{The Aitoff-Hammer projection of the 24 clusters initiated
with masses of $0.9\times 10^5 \msun$ (top two panels) and $0.9\times 10^4 \msun$ (bottom panel) in 
the $M_s=2$ infall model.  In the top panel any pixel with a particle in it is black,  a 0-1 grey scale.
All particles are equally visible in the top panel. 
In the middle and bottom panels particles beyond 16 kpc are weighted by their 
inverse distance squared.
In the middle panel the grey scale is 0-4 to emphasize the higher density regions, 
and in the bottom the grey scale is 0-40 which puts the middle and bottom panel on the
same stellar brightness scale.
}
\label{fig_ah45}
\end{figure}

Streams are very low sky density features which in photometric data alone become detectable 
through the use of color-magnitude diagram filtering \citep{Rockosi:02} which optimally weights some desired
common age, metallicity and distance population of stars in preference to field stars. Even
with such procedures currently available data generally lead to streams with a signal-to-noise per degree length
of stream of a few.  If the stream is at a sufficiently large distance such that the large
number of stars near the turnoff from the main sequence are not within the sample, 
about 15-20 kpc for current images, then the signal to noise drops further. The outcome is that finding star
streams is difficult at best and essentially impossible with current data to detect lower density regions of streams.

In Figure~\ref{fig_ah45} the
sky density is shown for all 24 streams of the bottom panel of  Figure~\ref{fig_xy45}. 
In the upper panel of Figure~\ref{fig_ah45} any pixel on the 
sky with a star is black, irrespective of the numbers. A pixel is 0.002 units in a the Aitoff-Hammer coordinate which
has a maximum of $2\sqrt{2}$. Scaling to 360\degr\ indicates that the pixels are approximately 0.13\degr\ on a side.
To more accurately represent sky density, in the middle panel the pixel density is  de-weighted with
the inverse distance squared for stars beyond 16 kpc and the resulting pixel density plotted with a grey scale
from 0 to 4 units. The bottom panel shows the expected sky densities for the $\sim 10^4\msun$ streams on a grey
scale of 0 to 40 to mimic the conversion to the brightness of stars.

The weighted map of the sky density of streams is a much more realistic view of what should be expected in
current sky maps \citep{Ngan:15}. The streams are thinner and shorter than they would be if all stars were detected, 
which will eventually be possible. However, the streams do not disappear entirely. 
That is, even though the length and width of streams is likely being under-estimated, 
the counts of streams is likely to be accurate for streams within double the color-magnitude main sequence turn-off detection distance. 
The details of the stream selection function will be studied more quantitatively in a future paper.

\begin{figure}
\begin{center}
\includegraphics[angle=-90,scale=0.8]{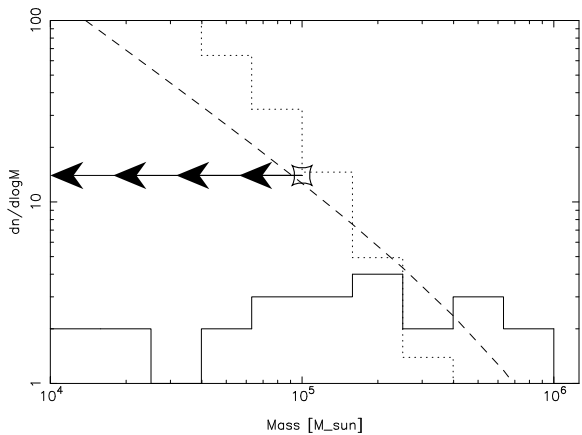}%{nGC.pdf}
\end{center}
\caption{The number of globular clusters as a function of mass in the PS1-$3\pi$ survey (solid
histogram) compared to the number of dense stellar systems expected to 
be formed in an $M^{-2} \exp{(-M/M_\ast)}$ distribution normalized to the top half dozen clusters
(dashed line). 
The dotted histogram is the cumulative
sum of the difference between the assumed initial distribution and the current numbers, 
which are expected to become stellar streams.
The pincushion symbol and left arrows is for the 14 streams recovered in the PS1-$3\pi$ area. 
The few stream masses known are in the few $\times 10^4\msun$ range. The streams
are put at $10^5\msun$ primarily on the basis that the heaviest ``lost" clusters should be the most visible star streams so
the stream mass should be as high as allowed.
}
\label{fig_ngc}
\end{figure}

\section{Stream - Cluster complementarity and Galaxy Assembly}

For every completely dissolved globular cluster, there will be a thin star stream with its stars 
distributed close to the orbit of the cluster. 
The streams also provide an indicator of when their progenitor cluster fell in through their length. A cluster
recently merged into the galaxy
will have a short tail and old additions should have very long tails. Creating a sample of both streams and globular clusters
with well-understood sampling is not yet possible, but an interesting start can be made.

The PS1-3$\pi$ survey provides nearly uniform sky coverage over a large sky area north of Declination -30.
The \citet{PS1} stream search is defined within a maximum distance of of 35 kpc, which here is taken
as a galactocentric radius. The minimum galactocentric radius is 
10 Kpc, inside of which streams are not found, possibly
because they encounter the stellar disk and its massive gas clouds and are quickly eroded  \citep{Amorisco:16}. 
Within the PS1-3$\pi$ sky \citet{PS1} 
recover 9 previously known thin streams (wider, dwarf galaxy streams are excluded here), and,
find 5 new ones for a total of 14. A few more streams are known in this sky area from other work
(compiled in \citet{GC:16}),  but PS1-3$\pi$ provides the
most uniform and large area of sky available at the moment.  A few of the streams have mass or luminosity estimates, 
such as GD-1 estimated to have a mass $2\times 10^4\msun$ \citep{KRH:10}. There is 
no estimate of how much mass associated with any stream may be missed at low surface brightness or
or dust obscuration. The simulations presented here indicate that the streams are likely
in the mass range around $10^5\msun$. Streams above this mass limit should be readily visible.

The \citet{Harris:96} catalog is used to construct a sample of globular clusters in the same sky volume.
Remnant halo clusters are those visible within the same sky volume as the star streams.
To a first approximation, the obscuration of the galactic disk should be about the same 
for streams and the halo population of globular clusters.
An upper limit on metallicity of [Fe/H] of -1 is imposed to focus on halo clusters,
although this only eliminates 3 very low luminosity clusters out of 27 in the PS1-3$\pi$ volume. 
The selected clusters have a mean galactocentric distance of 19 kpc.
Masses are assigned to the globular clusters using a uniform M/L of 2 solar units
and plotted in Figure~\ref{fig_ngc}. The pin cushion symbol shows the 14 PS1 streams, assigned 
a mass of $10^5\msun$, although that is likely an upper-limit, as indicated with the arrows.

The mass distribution of current epoch dense star clusters is a steep power law,
approximately $M^{-2}$ \citep{PZ:10}.
Since the high mass clusters in the halo are expected to have largely retained their initial mass, the
$M^{-2}$ can be fit to the higher mass clusters to predict the numbers at lower masses. 
In Figure~\ref{fig_ngc} the 
 dashed line is $7 (M/M_\ast)^{-1} \exp{(-M/M_\ast)}\Delta M/M$ with $M_\ast=10^6 \msun$ and 
 $\Delta \log_{10}{M}=0.2$,
which approximately fits the mass distribution of the half dozen clusters above $3\times 10^5 \msun$. 
The dotted histogram is the cumulative difference between the
clusters observed to be present and their assumed progenitor numbers, that is,
the number of ``lost" clusters. The difference between the
28.6 clusters expected from the dense star cluster progenitor distribution
 and the 14 clusters more massive than $10^5 \msun$ is 14.6. Given the uncertainty on the stream masses
this primarily means that the progenitor cluster distribution could not have been steeper
than $M^{-2}$.  It also suggests that essentially all of the massive streams in the PS1 volume have been found, but
there could be many more lower mass streams.

%\citep{GA:15}.

\section{Discussion}

The processes of galaxy assembly and star cluster  evaporation need to be included to provide
an accurate dynamical history of the development of stellar streams.
This paper develops a toy model
with a simple merger model and a cluster heating function that
approximates the internal dynamics of star clusters.
Clusters expand and cool as they lose mass, causing the stream linear density to vary with
distance from the progenitor clusters.
The density decline of streams in an evolving potential
contrasts with the effectively constant linear density
of streams produced
in static background potentials \citep{Kupper:12,Bovy:14,Carlberg:15a}, setting aside orbital phase variation.
The portions of the tidal streams created prior to merging of the satellite galaxy
into the host are often very widely dispersed.

The heated model clusters have mass loss  in rough agreement with detailed cluster dynamical studies. The
clusters  having a mass around $10^5\msun$  lose about 
one third of their mass in a Hubble time and $10^4\msun$ clusters largely evaporate in about 2/3 of a Hubble time.
The linear density of the resulting stellar streams declines about
a factor of ten between 10\degr\ and 50\degr\ from the progenitor cluster,
 although there is considerable case to case variation.
When projected onto the sky, the decline of the mean density away from the progenitor helps explain 
why the readily visible parts of most streams are relatively short, even though the simulated streams often
wrap a good fraction of their orbit around the galaxy. 
Dark matter sub-halos
create gaps along the streams, but because gaps are largely changes parallel to the orbital motion \citep{YJH:11},
the streams remains robust structures \citep{HW:99} and not readily blurred out. 
The result of these models is an improved conceptual understanding of
the development of streams and their sky density within an accreting halo, but
almost every aspect of the simulation can be improved in more
detailed future investigations.

The globular clusters around $10^5\msun$ that dissolve should all produce streams sufficiently bright 
to be detectable with optimized color-magnitude filtering methods if they are in an accessible part of the sky.
Therefore streams when combined with the remnant globular clusters in the same
volume can provide an estimate of the total cluster population present at early epochs. 
The current data do not have well determined stream masses and the stream selection function is not well understood. 
However, an interesting star can be made with the PS1-3$\pi$ survey 
which finds 14 streams \citep{PS1} in its volume. In the same sky volume there are 14 globular clusters to
$10^5\msun$. Making the assumption
that all the streams are all at or above $10^5 \msun$ in total mass then leads to the conclusion that there
were 28 clusters in in the progenitor cluster population to $10^5\msun$, which would be in accord with 
the expectations of an $M^{-2}$ initial cluster mass distribution. The minimum mass of the current stream sample is
almost certainly somewhat lower than $10^5\msun$ which then makes  $M^{-2}$ 
the upper limit to the shape of the progenitor mass distribution.
A possibly interesting implication is that if binary black
holes (BBH) at least partially form in clusters with low metallicity masses around 
$10^5\msun$ \citep{BBH:17} with the first few Gigayears prior
to assembly into the Milky Way, then a prediction of BBH numbers based on 
the current globular cluster population is an underestimate, by a factor of two or less, 
based on the stream counts.

\acknowledgements

This research was supported by CIFAR and NSERC Canada. An anonymous referee provided constructive criticism and
pointed to an error in the mass scaling in an earlier version.

\end{document}